\documentclass[aps,pra,twocolumn,superscriptaddress,showpacs]{revtex4}
\usepackage[dvips]{graphicx}
\usepackage{amssymb}
\usepackage{amsmath}

\begin{document}

%
\title{Long-path formation in a deformed microdisk laser}

\newcommand{\micron}{\mu\mbox{m}}

\newcommand{\NTTCS}{\affiliation{NTT Communication Science
    Laboratories, NTT Corporation, 2-4 Hikaridai, Seika-cho,
    Soraku-gun, Kyoto 619-0237, Japan}}
\newcommand{\KANAZAWA}{\affiliation{Faculty of Mechanical Engineering,
    Institute of Science and Engineering, Kanazawa University,
    Kakuma-machi, Kanazawa, Ishikawa 920-1192, Japan}}
\newcommand{\OKAYAMA}{\affiliation{Department of Information and
    Communication Engineering, Okayama Prefectural University, 111
    Kuboki, Soja, Okayama 719-1197, Japan}}
\newcommand{\WASEDA}{\affiliation{Department of Applied Physics,
    School of Advanced Science and Engineering, Waseda University,
    3-4-1 Okubo, Shinjuku-ku, Tokyo 169-8555, Japan}}
\author{Susumu Shinohara}\NTTCS
\author{Takehiro Fukushima}\OKAYAMA
\author{Satoshi Sunada}\KANAZAWA
\author{Takahisa Harayama}\WASEDA
\author{Kenichi Arai}\NTTCS
\date{\today}
\begin{abstract}
%
An asymmetric resonant cavity can be used to form a path that is much
longer than the cavity size.
We demonstrate this capability for a deformed microdisk equipped with
two linear waveguides, by constructing a multiply reflected periodic
orbit that is confined by total internal reflection within the
deformed microdisk and outcoupled by the two linear waveguides.
Resonant mode analysis reveals that the modes corresponding to the
periodic orbit are characterized by high quality factors.
From measured spectral and far-field data, we confirm that the
fabricated devices can form a path about 9.3 times longer than the
average diameter of the deformed microdisk.
\end{abstract}
%
\maketitle
%
\section{Introduction}
\label{introduction}
In the past two decades, asymmetric resonant cavities (ARCs) have
attracted considerable attention, because the introduction of asymmetry
has proven useful in generating directional emissions, while maintaining
high quality factors to some extent \cite{Noeckel97, Gmachl98}.
From the viewpoint of dynamical billiard theory, the introduction of
asymmetry generally accompanies the generation of ray-dynamical chaos
\cite{Chernov06}.
Thorough theoretical and experimental investigations elucidated the
relation between directional emission and ray-dynamical chaos
\cite{Schwefel04, Harayama11, Cao15}.

Although most of the previous work on ARCs has focused on directional
emission, ARCs have another noteworthy capability, namely that of
forming long optical paths by using multiple reflections at cavity
interfaces.
This capability was first demonstrated for a macroscopic (i.e.,
cm-sized) ARC with the aim of using it for gas sensing
\cite{Narimanov07, Qu08a, Qu08b}, where long paths are important for
increasing sensitivity.
In that work, a three-dimensional ARC with diameter of 5.24 cm was
fabricated of copper, and it was used to form a path of up to 6 m.
The idea of utilizing multiple reflections to form a long path can be
found in relation to traditional gas sensing cavities consisting of
two facing mirrors such as White cells and Herriott cells
\cite{White42, Herriott65}.

We brought a similar idea to the micro-scale, and achieved a 2.79 mm
path for a microcavity with an average diameter of 300 $\micron$
\cite{Shinohara14}, with the aim of using the path to realize a
compact laser chaos device.
Recently, laser chaos has attracted renewed interest, because of its
usefulness as an entropy source for physical random number generation
\cite{Uchida12}.
Laser chaos with a GHz bandwidth can be easily generated by a
semiconductor laser with an external cavity for delayed optical
feedback \cite{Ohtsubo08}.
The external cavity must be at least a few mm long in order to obtain
laser chaos suitable for random number generation.
This requirement imposes a bottleneck on the device size, if we
restrict ourselves to using a one-dimensional external cavity.
However, a two-dimensional external cavity with a sufficiently long
path makes it possible to realize an ultra-small integrated laser
chaos device whose footprint can be less than 1 mm$^2$
\cite{Sunada14}.

\begin{figure}[t]
\includegraphics[width=0.45\textwidth]{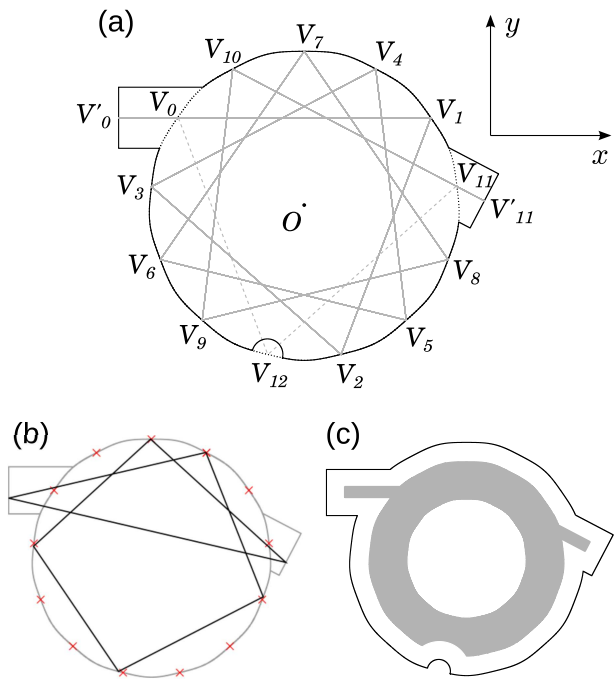}
\caption{\label{fig:cavity} (Color online) (a) Deformed microdisk with
  two linear waveguides attached at $V_0$ and $V_{11}$ and a
  half-circular scatterer at $V_{12}$. (b) The stable periodic orbit
  that appears when the scatterer at $V_{12}$ is absent. (c) The
  contact area (gray region), where currents are injected.}
\end{figure}

Ref. \cite{Shinohara14} introduced a new type of cavity consisting of
a deformed microdisk and two linear waveguides.
This cavity enables the formation of a path much longer than the
cavity diameter (see Fig. \ref{fig:cavity} (a) for the cavity shape).
In contrast to a one-dimensional waveguide, long-path formation in
two-dimensional slab cavity is highly nontrivial, because there is no
transverse confinement, and only focusing at the cavity boundary
prevents light from diffusing.
In Ref. \cite{Shinohara14}, the path formation was experimentally
confirmed by injecting light from one of the linear waveguides and
measuring the output from the other waveguide.

In this paper, we provide more detailed evidence for the path
formation by performing systematic wave calculations for the cavity
and carrying out experiments that allow direct theory-experiment
comparisons.
We also fabricate a relatively small device with an average diameter
of 60 $\micron$ (note that the average diameter of the device in
Ref. \cite{Shinohara14} is 300 $\micron$), whose experimental results
can be directly compared with those of a resonant mode analysis.
Firstly, we show that the resonant modes corresponding to the long
path constitute a dominant group of high quality factor modes,
implying that the formed path is well confined.
By comparing results from the resonant mode analysis and those from
Gaussian optical theory, we discuss the deviations from an ideal
Gaussian beam caused by the presence of ray chaos.
Secondly, we show that experimentally measured far-field patterns can
be well reproduced by that of a resonant mode corresponding to the
long path.
Moreover, we reveal that measured modal spacings in optical spectra
show excellent agreement with the theoretical estimates assuming the
long path.
All these experimental results provide decisive evidence for the
formation of a path that is about 9.3 times longer than the average
diameter of a deformed microdisk.
Additionally, we report that single-mode lasing occurs for continuous
wave pumping.

This paper is organized as follows: In Sec. \ref{sec:cavity}, we
introduce our cavity design and present a stability analysis for a
long periodic orbit. In Sec. \ref{sec:modes}, we describe resonant
mode analysis with particular attention to modes corresponding to a
long periodic orbit. In Sec. \ref{sec:experiments}, we provide
experimental data for far-field and spectral measurements that confirm
the existence of the designed long path. Sec. \ref{sec:summary}
consists of a summary and discussion.

\section{Cavity design}
\label{sec:cavity}
We consider a cavity consisting of a deformed disk and two linear
waveguides as shown in Fig. \ref{fig:cavity} (a).
The shape of the deformed disk is defined by
\begin{equation}
r(\phi)=r_0\left[1-\varepsilon\sin(Q\phi)\right],
\label{eq:cavity}
\end{equation}
where $(r,\phi)$ are the polar coordinates, and $r_0$ and $\varepsilon$
are the size and deformation parameters, respectively.
In this paper, we fix $Q$ = 13, but the idea described below can apply
to the other integer $Q$, provided that the corresponding cavity has a
star-polygonal periodic orbit.
When we employ the shape given by Eq. (\ref{eq:cavity}) as the
boundary for a dynamical billiard, we have a star-polygonal periodic
orbit with 13 vertices labeled $V_0, V_1, \cdots, V_{12}$, as shown in
Fig. \ref{fig:cavity} (a).
These vertices correspond to the minimum curvature points of the
boundary, and are defined by the polar angles $\phi_m$ $=$
$\pi(17-4m)/26$ $(m=0,1,\cdots,12)$.
As the star-polygonal orbit is traced until it closes, we have a $P$
$=$ $4$ times rotation around the origin $O$.
In other words, the winding number of the star-polygonal orbit is
$P/Q$ $=$ $4/13$.
A linear stability analysis reveals that the periodic orbit is stable
for $0\leq \varepsilon \lesssim 0.00588$, and it is confined by total
internal reflection when the refractive index of the cavity exceeds
1.76 (when we assume that the refractive index outside the cavity is
1).

\begin{figure}[t]
\includegraphics[width=0.5\textwidth]{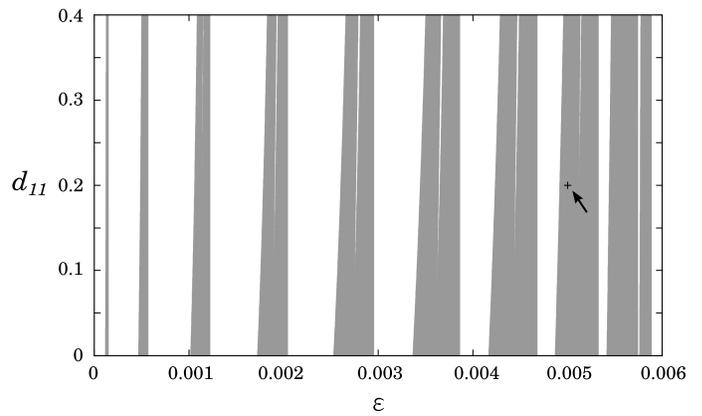}
\caption{\label{fig:diagram} Stability diagram for the
  open-star-polygonal periodic orbit, where $\varepsilon$ is the
  deformation parameter and $d_{11}$ is the scaled length of the
  waveguide at $V_{11}$, i.e., $d_{11}$ $=$
  $\overline{V_{11}V'_{11}}/r_0$ (note that the scaled length of the
  waveguide at $V_{0}$ is fixed at $d_0=1.90565\times d_{11}$). The
  open-star-polygonal periodic orbit is linearly stable in the gray
  regions (i.e., $|\mbox{Tr}\,M|<2$ with $M$ the monodromy matrix). In
  this work, the parameter values are fixed at
  $(\varepsilon,d_{11})=(0.005,0.2)$ (indicated by an arrow).}
\end{figure}

At the vertex $V_0$ (resp. $V_{11}$), we attach a linear waveguide
parallel to a periodic orbit segment $V_0 V_1$ (resp. $V_{10}
V_{11}$).
This eliminates the star-polygonal periodic orbit.
Instead, we have a self-retracing periodic orbit connecting both
waveguide ends $V_0'$ and $V_{11}'$.
This orbit includes the entire star-polygonal orbit except for the two
line segments $\overline{V_0 V_{12}}$ and $\overline{V_{11} V_{12}}$.
It is this open-star-polygonal periodic orbit that we focus on in this
paper.
To prevent the other periodic orbits from appearing, we place a
half-circular scatterer at the vertex $V_{12}$, as shown in
Fig. \ref{fig:cavity} (a).
Without this scatterer, there is a stable periodic orbit as shown in
Fig. \ref{fig:cavity} (b).
The (one-way) path length $L_{*}r_0$ of the open-star-polygonal
periodic orbit is given by
\begin{equation}
L_* r_0=\left[
2N(1-\varepsilon)\sin\left(\pi{\displaystyle \frac{P}{Q}}\right)+d_0+d_{N}
\right]r_0,
\end{equation}
where $N$ is the index for the vertex with the waveguide (we assume
that the other waveguide is attached to the vertex $V_0$).
For the open-star-polygonal orbit shown in Fig. \ref{fig:cavity} (a),
we have $L_{*} r_0=18.596$ $\times$ $r_0$, that is, the path is about
9.3 times longer than the average cavity diameter.

\begin{figure}[b]
\includegraphics[width=0.5\textwidth]{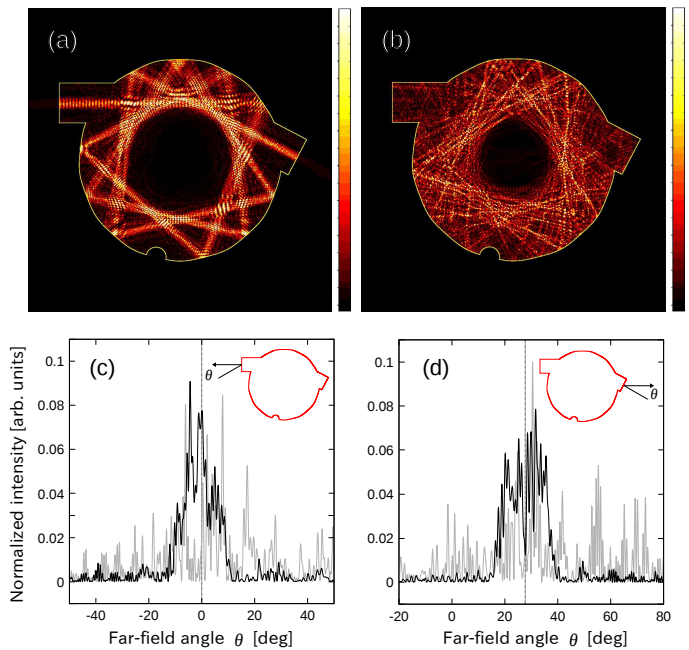}
\caption{\label{fig:wfunc} (Color online) (a) and (b) show the
  intensity distributions of the wave functions for the resonances:
  (a) $k r_0$ $=$ 224.165 $-$ $i0.011$ and (b) $k r_0$ $=$ 224.430 $-$
  $i0.010$, where a lighter color indicates a higher intensity. (c)
  and (d), respectively, show far-field emission patterns for the mode
  shown in Fig. \ref{fig:wfunc} (a) (black curves) and for the mode
  shown in Fig. \ref{fig:wfunc} (b) (gray curves).}
\end{figure}

\begin{figure}[t]
\includegraphics[width=0.4\textwidth]{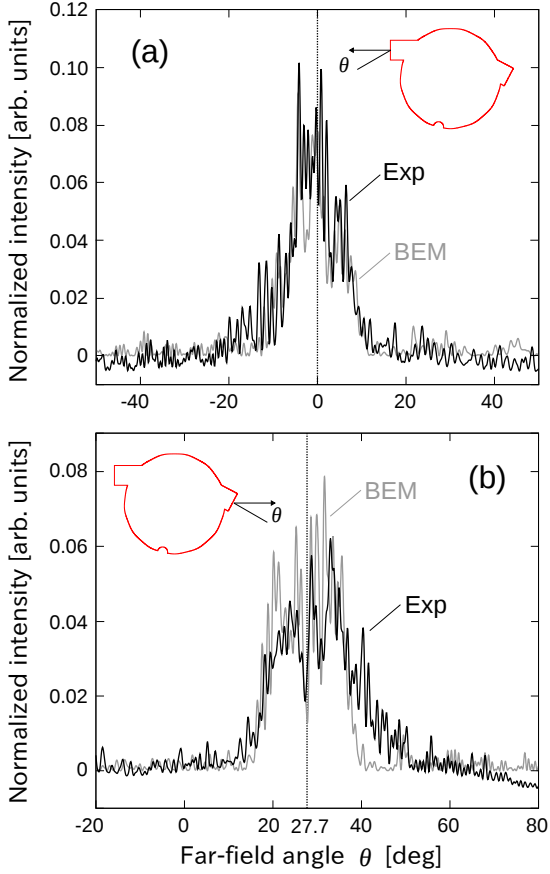}
\caption{\label{fig:ffp1} (Color online) Far-field patterns
  (normalized) for a device with $r_0=30\,\micron$: (a) emission to
  the left side of the cavity, (b) emission to the right side. The
  direction of the waveguide is $\theta$ $=$ 0 degrees in (a), while
  it is $\theta=27.7$ degrees in (b). Experimental data for a pumping
  current of 90 mA are plotted with black curves, while numerical data
  obtained with the BEM calculation are plotted with gray curves.}
\end{figure}

\begin{figure}[t]
\includegraphics[width=0.4\textwidth]{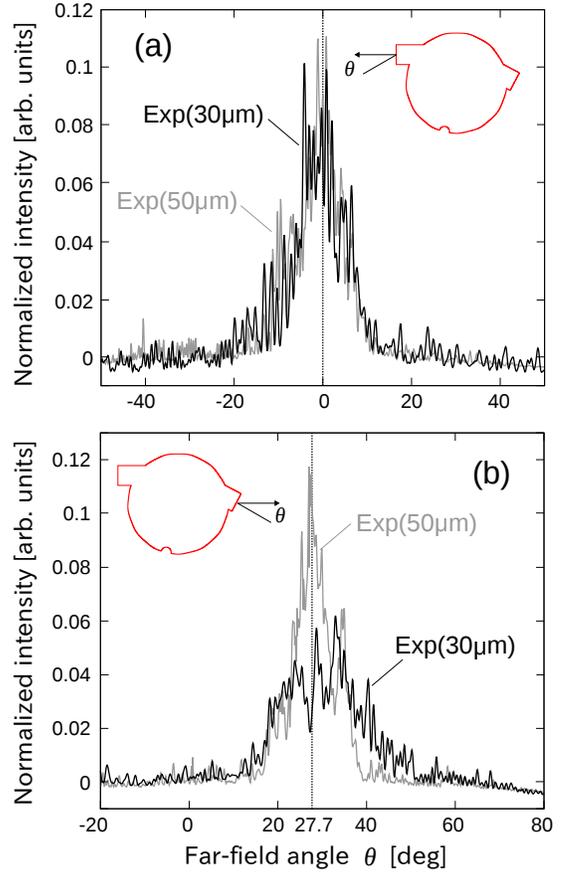}
\caption{\label{fig:ffp2} (Color online) Far-field patterns
  (normalized) for devices with $r_0=30\,\micron$ and
  $r_0=50\,\micron$: (a) emission to the left side of the cavity, (b)
  emission to the right side. The direction of the waveguide is
  $\theta$ $=$ 0 degrees in (a), while it is $\theta=27.7$ degrees in
  (b). Experimental data for $r_0=30\,\micron$ are plotted with black
  curves, while those for $r_0=50\,\micron$ are plotted with gray
  curves. The pumping current is 90 mA in all cases.}
\end{figure}

The linear stability of the open-star-polygonal periodic orbit depends
on three parameters, namely the deformation parameter $\varepsilon$
and the lengths of the two linear waveguides.
%
For the waveguide attached to $V_0$ (resp. $V_{11}$), we use $d_{0}$
(resp. $d_{11}$) to denote the length scaled by $r_0$, i.e.,
$d_i=\overline{V_i V'_i}/r_0$ $(i=0,11)$.
In what follows, we fix the ratio $d_{0}/d_{11}$ $=$ $1.90565$, and
consider $d_{11}$ to be a free parameter, in addition to
$\varepsilon$.
The asymmetry of the waveguide lengths breaks the otherwise existing
mirror symmetry of the cavity (this results in asymmetric emission
patterns as seen in Figs. \ref{fig:wfunc}, \ref{fig:ffp1} and
\ref{fig:ffp2}).

We evaluated the monodromy matrix $M$ for the open-star-polygonal
periodic orbit numerically for each set of $\varepsilon$ and $d_{11}$,
and obtained the stability diagram shown in Fig. \ref{fig:diagram},
where the periodic orbit is stable (i.e., $|\mbox{Tr}(M)|<2$) in the
gray regions.
We hereafter fix the parameter values at $(\varepsilon,d_{11})$ $=$
$(0.005,0.2)$ as indicated by the arrow in Fig. \ref{fig:diagram}.
We also fix the waveguide widths at $W_0=W_{11}=0.395875\times r_0$,
so that they are much larger than the beam spot sizes at the waveguide
ends.

\begin{figure}[b]
\includegraphics[width=0.5\textwidth]{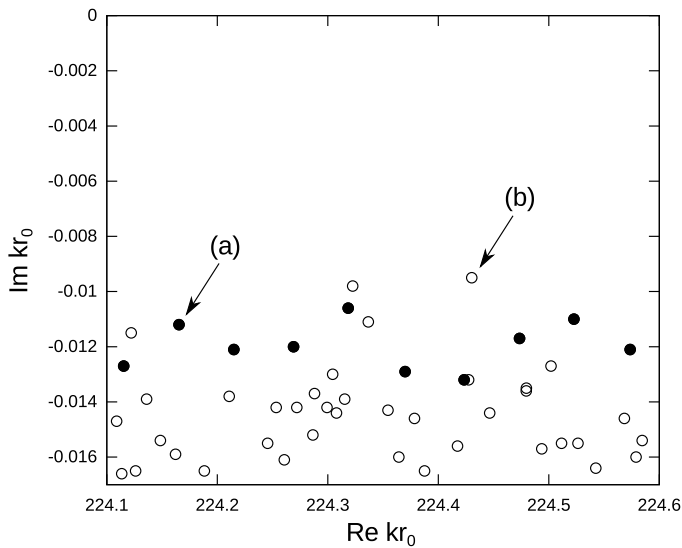}
\caption{\label{fig:resonances} The distribution of the resonances,
  where the star modes are plotted with filled circles
  ($\bullet$). The star modes appear regularly with the average mode
  spacing $\Delta k r_0$ = 0.0512. The wave functions corresponding to
  (a) and (b) are shown in Figs. \ref{fig:wfunc} (a) and
  \ref{fig:wfunc} (b), respectively.}
\end{figure}

\section{Resonant mode analysis}
\label{sec:modes}
We performed a numerical analysis of the resonant modes for the cavity
defined in Fig. \ref{fig:cavity} (a) by using the boundary element
method (BEM) \cite{Wiersig03}.
The resonant modes are the solutions of the Helmholtz equation
$[\nabla^2$ $+$ $n(x,y)^2 k^2]$ $\psi(x,y)=0$, where $k$ is the
wavenumber and $n(x,y)$ is the effective refractive index.
We set $n(x,y)$ $\equiv$ $3.3$ (GaAs) inside the cavity, while
$n(x,y)$ $\equiv$ $1$ (air) outside the cavity.
As described later, we fabricated the cavity with an unstrained GaAs
single-quantum-well structure.
Since the emissions from these lasers are usually TE-polarized, we
employ the cavity interface conditions for TE-polarization (i.e., both
$\psi$ and $(1/n^2)\partial\psi/\partial n$ are continuous at the
cavity interface, where $\partial/\partial n$ is the derivative normal
to the boundary).
For the TE-polarization, the wave function $\psi$ represents the
$z$-component of the magnetic field vector, i.e., $H_z=\mbox{Re}[\psi
  e^{-ickt}]$.
At infinity, we assume the outgoing wave condition $\psi\propto
e^{ikr}/\sqrt{r}$, which yields complex wavenumbers $k$ with
$\mbox{Im}\,k<0$.

Figure \ref{fig:resonances} shows the distribution of (dimensionless)
complex wavenumbers $k r_0$ for resonant modes found numerically
around $\mbox{Re}\,k r_0=224.4$.
This $\mbox{Re}\,k r_0$ value corresponds to the cavity size
$r_0=30\,\micron$ when the wavelength is $\lambda=0.84\,\micron$,
which is one of our experimental conditions as discussed later.

According to Gaussian optical theory \cite{Tureci02}, Gaussian beam
modes that are localized along the open-star-polygonal orbit are
expected to exist when $\mbox{Re}\,k r_0$ values are sufficiently
large.
We call these modes {\sl star modes}, and we detected many of them in
the BEM calculation (the star modes were clearly identified by their
strong localization along the open-star-polygonal orbit).
In Fig. \ref{fig:resonances}, the star modes are indicated by filled
circles.
We can see that the star modes constitute a dominant group among the
high quality factor modes.
For the star mode labeled (a) in Fig. \ref{fig:resonances} (i.e., $k
r_0$ $=$ 224.165 $-$ $i$ 0.011), we show the corresponding spatial
intensity pattern of the wave function, $|\psi(x,y)|^2$, in
Fig. \ref{fig:wfunc} (a), where we can see strong localization along
the open-star-polygonal orbit.
The far-field pattern of this mode is shown in Figs. \ref{fig:wfunc}
(c) and \ref{fig:wfunc} (d) (black curves).
Figure \ref{fig:wfunc} (c) shows the emission to the left side of the
cavity (the side of $V'_{0}$), while Fig. \ref{fig:wfunc} (d) shows
that to the right side (the side of $V'_{11}$).
We can confirm that there is a bi-directional emission that is in good
agreement with the directions of the two attached linear waveguides
(i.e., $\theta=0$ degrees in Fig. \ref{fig:wfunc} (c), while
$\theta=27.7$ degrees in Fig. \ref{fig:wfunc} (d)).
Note that the far-field patterns are normalized.

Unexpectedly, we found another type of high quality factor mode in the
resonance distribution (labeled (b) in Fig. \ref{fig:resonances}).
Its wave function is shown in Fig. \ref{fig:wfunc} (b), and its
far-field patterns are shown in Figs. \ref{fig:wfunc} (c) and
\ref{fig:wfunc} (d) (gray curves).
A comparison between polygonal periodic orbits and the wave function
pattern did not reveal a single dominant periodic orbit that explains
the wave function pattern.
However, the wave function appears to be localized along straight ray
segments, especially those of unstable triangular (three-bounce)
orbits.
Such a mode might be understood as the scar mode \cite{Heller84}.
This localization is interesting in itself, and deserves further
investigation, but this will constitute future work, since our main
focus here is on the star modes.

Gaussian optical theory \cite{Tureci02} states that the beam waist
spot size and position are directly related to the stability of the
periodic orbit.
For the open-star-polygonal orbit, the waveguide ends $V'_0$ and
$V'_{11}$ are the beam waist positions, and their spot sizes are
estimated as
\begin{equation}
w=\tilde{w} \frac{r_0}{\sqrt{k r_0}},
\end{equation}
where $\tilde{w}$ $=$ $0.961$ for the beam waist at $V'_0$, while
$\tilde{w}$ $=$ $0.898$ for the beam waist at $V'_{11}$.
The beam divergence $\theta_d$ in the far field can be estimated by
the formula $\theta_d\approx \lambda/(\pi w)=2/(\tilde{w}\sqrt{k
  r_0})$, where $w$ is the beam spot size.
This formula yields $\theta_d$ $=$ $8.0$ degrees for the emission at
$V'_0$, and $\theta_d=8.5$ degrees for the emission at $V'_{11}$.
On the other hand, for the far-field data obtained with the BEM
calculation, we found that $\theta_d$ $=$ $11.8$ degrees for the
emission at $V'_0$ [black curve in Fig. \ref{fig:wfunc} (c)], while
$\theta_d$ $=$ $14.7$ degrees for the emission at $V'_{11}$ [black
  curve in Fig. \ref{fig:wfunc} (d)].
We consider that this broadening of the peaks is mainly caused by the
deformation of the Gaussian beams due to the finiteness of the
wavelength and the existence of ray chaos.

For the 10 star modes that we detected in our BEM calculations, we
found an almost regular modal spacing that fluctuated from $0.0492$ to
$0.0540$.
We consider this relatively large fluctuation to be related to {\sl
  dynamical tunneling} \cite{Davis81}, which is known generally to
occur in ray-chaotic microcavities \cite{Shinohara10, Yang10, Song12},
resulting in some portion of the intensity being leaked to ray chaotic
orbits and spreading throughout the cavity.
Nevertheless, we found that the average modal spacing is $\Delta k
r_0=0.0511$, which closely agrees with a theoretical estimate based
on the optical path length, i.e., $\Delta k r_0=\pi/(n L_{*})=0.0512$,
where $L_{*}$ $=$ $18.596$ is the (one-way) path length of the
open-star-polygonal orbit normalized by $r_0$.

As for the decay rates, the average value of $\mbox{Im}\,k r_0$ for
the 10 star modes is $-0.0120$.
In the short-wavelength limit, $\mbox{Im}\,k r_0$ can be approximately
expressed as \cite{Tureci02}
\begin{equation}
\mbox{Im}\,k r_0=\frac{\ln\left({1/\cal R}\right)}{2n L_{*}},
\label{eq:decay_rate}
\end{equation}
where ${\cal R}=0.286$ is the Fresnel reflection coefficient for
normal incidence and TE-polarization.
This formula yields $\mbox{Im}\,k r_0=-0.0102$.
The average decay rate of the actual star modes is about 20\% larger
than this theoretical estimate.
This effect can also be considered to result from the dynamical
tunneling.
That is, there is an additional channel for the leakage, or an
emission that is formed by the dynamical tunneling \cite{Shinohara10}.


\begin{figure}
\includegraphics[width=0.4\textwidth]{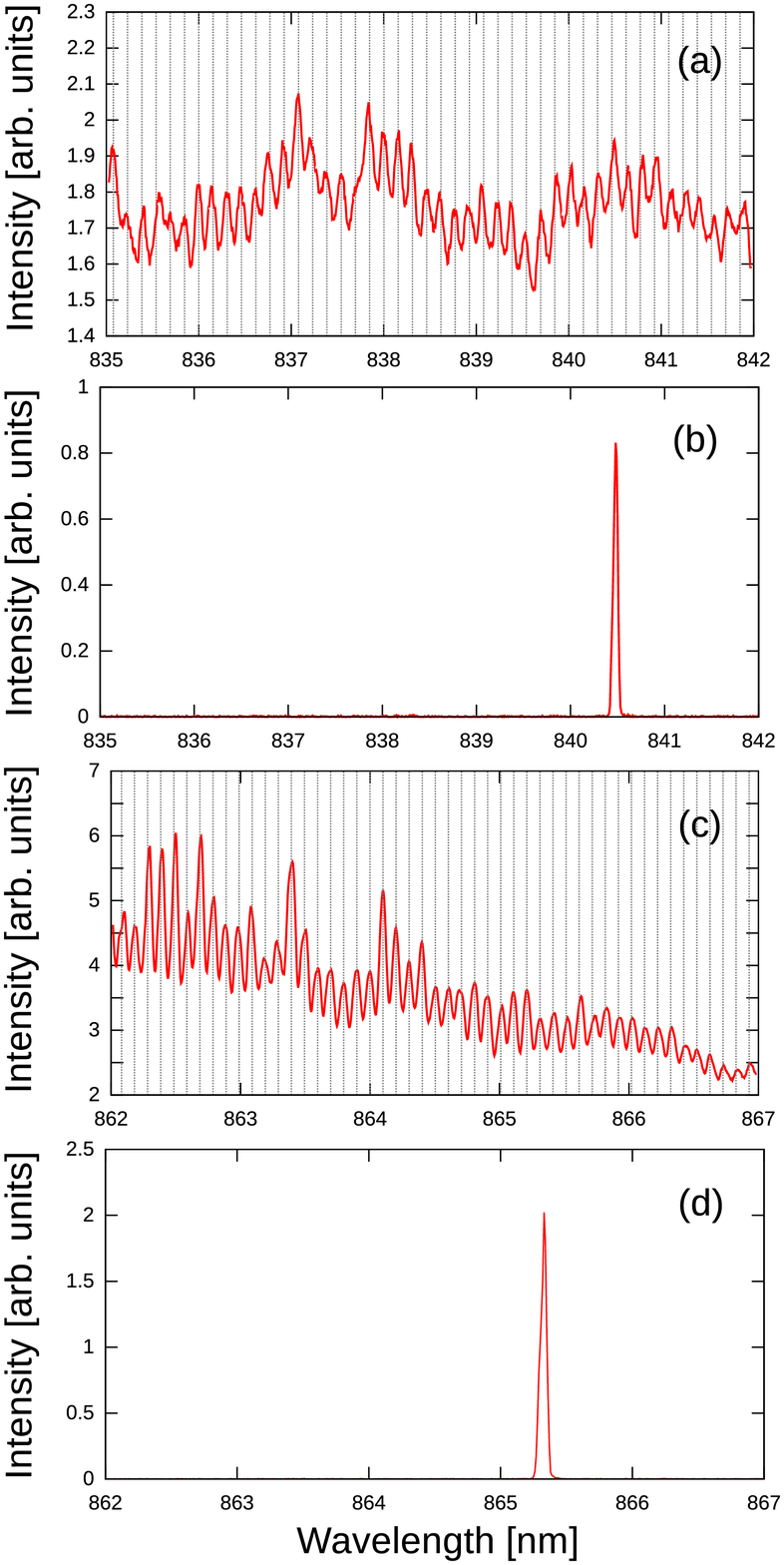}
\caption{\label{fig:spectra} (Color online) Optical spectra: (a)
  $r_0=30\,\micron$ and below the threshold, (b) $r_0=30\,\micron$ at
  90 mA, (c) $r_0=50\,\micron$ and below the threshold (d)
  $r_0=50\,\micron$ at 90 mA. In (a) and (c), respectively,
  equidistant vertical dotted lines with 0.154 nm and 0.101 nm
  spacings are plotted as an eye guide.}
\end{figure}

\section{Experimental data}
\label{sec:experiments}
We fabricated GaAs single-quantum-well (SQW) microcavity lasers whose
cavity shape is defined in Fig. \ref{fig:cavity} (a) with average
diameters of $r_0$ $=$ $30\,\micron$ and $r_0$ $=$ $50\,\micron$.
The GaAs SQW heterostructures were grown by metal-organic
chemical-vapor deposition (MOCVD) on an n-type GaAs substrate with a
1.5-$\micron$ $n$-Al$_{0.5}$Ga$_{0.5}$As lower cladding layer, a
0.2-$\micron$ $n$-Al$_x$Ga$_{1-x}$As ($x=0.5-0.2$) graded region, a
10-nm GaAs quantum well layer, a 0.2-$\micron$
$p$-Al$_x$Ga$_{1-x}$As($x=0.2-0.5$) graded region, a 1.5-$\micron$
$p$-Al$_{0.5}$Ga$_{0.5}$As upper cladding layer, a 0.2-$\micron$
$p$-GaAs contact layer, and a 400-nm SiO$_2$ layer.
The etching depth from the top of the contact layer was about 4.2
$\micron$, which was 2.3 $\micron$ below the active layer.
The contact area was etched through the SiO$_2$ layer, and the
$p$-electrode metal was formed over the contact area and part of the
surrounding SiO$_2$ layer using a liftoff process (see
Ref. \cite{Fukushima04} for details of the fabrication process).
Figure \ref{fig:cavity} (c) shows the contact area shape, which was
designed to fit to the star modes.
It mainly consists of an annular area, whose outer and inner
boundaries have the shape defined by Eq. (\ref{eq:cavity}) with the
average radii $r_0-5\,\micron$ and $r_0/2$, respectively, where $r_0$
$=$ $30\,\micron$ or $50\,\micron$.
In addition, along the two linear waveguides, $2\,\micron$-width
linear contact areas are formed.
The $5\,\micron$ margin for the outer boundary of the contact area was
necessary for the liftoff process used for forming the $p$-electrode
metal \cite{Fukushima04}.
The lasers were electrically pumped by current injection with
continuous wave (CW) pumping at room temperature.
The lasing threshold was 30 mA for the device with $r_0$ $=$
$30\,\micron$, and 37 mA for the device with $r_0$ $=$ $50\,\micron$.
This implies that the threshold current density is 2.4 kA/cm$^2$ for
$r_0$ $=$ $30\,\micron$, and 0.84 kA/cm$^2$ for $r_0$ $=$
$50\,\micron$ (the contact area is approximately given by
$\pi[(r_0-5)^2-r_0^2/4]$).
We consider that the higher threshold current density for $r_0$ $=$
$30\,\micron$ is due to the threshold increase by a temperature rise
under CW pumping, which is expected to be more significant for a
smaller cavity.

In our experiments, we observed bi-directional emissions in good
agreement with the directions of the two attached waveguides.
Figure \ref{fig:ffp1} (black curves) shows measured far-field patterns
for the device with $r_0$ $=$ $30\,\micron$ when the pumping current
was 90 mA.
The far-field patterns were measured by rotating the photodetector
with a slit around the device.
The slit width was 0.5 mm and the radius of the rotation was 30 cm,
which yielded a resolution of 0.1 degrees.
As the result of the Lloyd's mirror effect, the direct output from the
active layer and the output once reflected at the GaAs substrate
produced an interference pattern on the plane vertical to the
horizontal plane where the device was placed.
The height of the photodetector was adjusted to the position of the
first-order interference to maximize the detected intensity.

The uniform background radiation caused by spontaneous emission is
subtracted from the experimental data in Fig. \ref{fig:ffp1}, and the
far-field patterns are normalized.
When we compare these results with the numerical results of the BEM
calculation superposed in Fig. \ref{fig:ffp1} (gray curves), we can
confirm good agreement between the experimental and theoretical data.
Note that the condition for the BEM calculation was the same as that
for the experiment (i.e., $r_0=30\,\micron$).
Figure \ref{fig:ffp2} shows measured far-field patterns for the device
with $r_0$ $=$ $50\,\micron$ (gray curves) together with the results
for the device with $r_0$ $=$ $30\,\micron$ (black curves).
We again see good agreement with the theoretical prediction, as well
as a slight far-field peak narrowing for $r_0$ $=$ $50\,\micron$,
especially in Fig. \ref{fig:ffp2} (b).
This is consistent with the fact that the beam divergence $\theta_d$
is inversely proportional to $\sqrt{k r_0}$.

Figure \ref{fig:spectra} (a) shows the optical spectrum obtained for
the device with $r_0=30\,\micron$ when pumped below the threshold.
In the spectral measurement, the photodetector was placed in the
direction of the left linear waveguide.
We interpret the peaks we observed here as the cavity modes coupled
with the amplified spontaneous emission.
We can see a regular modal spacing $\Delta\lambda$ $=$
$0.154\,\mbox{nm}$.
This value is in excellent agreement with the theoretical estimate of
0.153 nm, which takes account of the dispersion, i.e., $\Delta\lambda$
$=$ $\lambda^2/[2nL_{*}r_0(1-(\lambda/n)(dn/d\lambda)]$, where
$\lambda$ $=$ $0.84\,\micron$ is the lasing wavelength, $L_{*}r_0$ $=$
$557.89\,\micron$ is the (one-way) path length of the
open-star-polygonal orbit, and $dn/d\lambda$ $=$ $-1.0\,\micron^{-1}$
is the dispersion \cite{Casey78}.
Figure \ref{fig:spectra} (c) shows an optical spectrum for the device
with $r_0=50\,\micron$ when pumped below the threshold.
The measured modal spacing is $\Delta\lambda$ $=$ $0.101\,\mbox{nm}$,
while the theoretical estimate based on the path length is
$\Delta\lambda$ $=$ $0.097\,\mbox{nm}$.

From the above far-field and spectral data, we concluded that the star
modes existed and were successfully excited in our fabricated devices.
As for the unexpected high quality factor modes that we observed in
the BEM calculation (i.e., the modes exemplified in
Fig. \ref{fig:wfunc} (b)), their existence cannot be confirmed in the
spectra shown in Figs. \ref{fig:spectra} (a) and \ref{fig:spectra}
(c).
This might be due to our contact pattern [Fig. \ref{fig:cavity} (c)],
which we designed to preferentially excite the star modes.
Also, as can be seen in the spatial intensity pattern of the
unexpected high quality factor mode shown in Fig. \ref{fig:wfunc} (b),
it does not couple significantly with the linear waveguides.
Thus, the photo detector placed in the direction of the waveguide
might not be able to capture the emission from an unexpected high
quality factor mode, even if it is excited.

Interestingly, when the devices are pumped above the threshold, we
observed single-mode lasing, as shown in Fig. \ref{fig:spectra} (b)
for the device with $r_0$ $=$ $30\,\micron$ and in
Fig. \ref{fig:spectra} (d) for the device with $r_0$ $=$
$50\,\micron$, where the pumping currents were 90 mA in both cases.
For chaotic cavity lasers, single-mode lasing has been experimentally
observed with CW pumping \cite{Audet07, Kim12,
  Sunada13, Sunada16}.
For chaotic cavity lasers, it has been numerically observed that
nonlinear interactions among modes reduce the number of lasing modes
\cite{Sunada05}, and experimentally observed single-mode lasing is
attributed to large modal overlaps between resonant modes
\cite{Sunada16}.

The excitation of multiple longitudinal Gaussian beam modes has been
observed with CW pumping for a stable Fabry-Perot cavity laser
\cite{Fukushima12}.
This is in contrast to our results.
We consider this difference to be caused by the fact that our Gaussian
beam modes are affected by dynamical tunneling.
As discussed in Sec. \ref{sec:modes}, the intensity pattern of our
Gaussian beam mode was slightly deformed from an exact Gaussian beam
because of the intensity leakage throughout the cavity.
This can result in a fluctuation in the decay rates (i.e.,
$\mbox{Im}\,kr_0$) as well as an increase in the modal overlaps.
The former makes a certain mode preferential for lasing, while the
latter can cause the suppression of the other lasing modes.
We expect these effects to lead to single-mode lasing, although more
thorough investigations are needed for a definitive understanding.

\begin{figure}[t]
\includegraphics[width=0.4\textwidth]{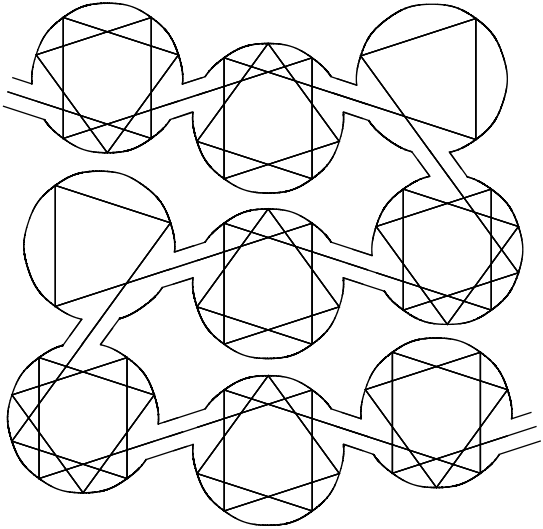}
\caption{\label{fig:cells} Long path formation by coupling 9 identical
  deformed disks with $Q=10$ and an average radius $r_0$. The total
  path length is about 107 $\times$ $r_0$, whereas the entire system
  length is about 6 $\times$ $r_0$.}
\end{figure}

\section{Summary and Discussion}
\label{sec:summary}
We studied both theoretically and experimentally a deformed
microdisk designed for long path formation.
We performed resonant mode analysis to reveal that the long-path modes
were characterized by high quality factors.
For fabricated deformed microdisks with average radii $r_0$
$=$ 30 $\micron$ and 50 $\micron$, we presented measured far-field and
spectral data that confirmed the path length of $18.596\times r_0$.

Based on this approach, the path length can be in principle extended
by increasing the size parameter $r_0$ or the number of vertices $Q$,
although the finite wavelength limits the maximum $Q$ value, since
each vertex needs to be resolved by the light wave so that a path is
distinctly formed.
Another interesting way of extending a path is to couple many cavities
as illustrated in Fig. \ref{fig:cells}, where 9 identical deformed
disks with $Q=10$ and an average radius $r_0$ are connected.
In this example, the total path length adds up to about 107 $\times$
$r_0$, whereas the entire system length is about 6 $\times$ $r_0$.
We performed a linear stability analysis as described in
Sec. \ref{sec:cavity} for this self-retracing periodic orbit, and
confirmed the existence of cavity parameter regimes where the
periodic orbit becomes stable.
In this way, we can use two dimensions efficiently to obtain a long
path.
Conventionally, a spiral geometry has been adopted for long path
formation \cite{Lee12}.
For a spiral waveguide, bending losses are inevitable, limiting the
maximum curvature of a spiral and thus the entire system size.
Our approach may be useful as an alternative that is free from bending
losses.

The cavity studied here can also be viewed as the one exhibiting
relatively good emission directionality.
Bi-directionality is inevitable for our cavity design based on the
open-star-polygonal periodic orbit.
However, we believe that unidirectionality can be achieved, by letting
one of the waveguide facets coincide with a cleavage facet, and
coating it with high reflectivity film.

%

\appendix

\end{document}